\documentclass[11pt]{article}
\title{Symbolic Dynamics of Homoclinic Orbits in a Symmetric Map}
\begin{document}
\author{Zai-Qiao Bai  and Wei-Mou Zheng\\
 Institute of Theoretical Physics, Academia Sinica, Beijing
 100080, China\\
 }
 \maketitle
 \baselineskip 18pt \vskip 5mm
 \begin {minipage}{140mm}
 \begin {center} {\bf Abstract} \end {center}
 \baselineskip 18pt

 Symbolic dynamics for homoclinic orbits in the
 two-dimensional symmetric map,
 $x_{n+1}+cx_{n}+x_{n-1}=3x_{n}^3$, is discussed.
 Above a critical $c^{\ast}$,
 the system exhibits a fully-developed horse-shoe so
 that its global behavior is described by a
 complete ternary symbolic dynamics.
 The relative location of
 homoclinic orbits is determined by their
 sequences according to a simple rule, which can be used to
 numerically locate orbits in phase space.
 With the decrease of $c$, more and more pairs of
 homoclinic orbits collide and disappear.
 Forbidden zone in the symbolic
 space induced by the collision is discussed.

 \end {minipage}
 \newpage
 \section {Introduction}
 This paper concerns locating homoclinic orbits (HOs) for maps.
 The topic received some recent interests because it is
 closely related to finding spatially localized oscillatory
 solutions in one-dimensional lattices (see \cite{Rep} and
 references therein). Numerical
 methods usually involve a shooting strategy or an
 assumption of the initial topology of HOs \cite {Fried,Far,Bey1,Bey2}.
 In both cases, a deep understanding of the qualitative
 behavior of maps is important.
 As a coarse-grained description of dynamics, symbolic
 dynamics(SD) has been productively applied to one- and
 two-dimensional maps\cite{Zheng}. The effectiveness of SD
 relies on the hyperbolic nature of the underlying
 dynamics. Therefore, SD should be a powerful tool to
 describe the qualitative behavior of HOs when
 the system is dominated by unstable motion. In this paper,
 we shall discuss this through a specific map.

 The system we consider is the map
 with its orbits $\{x_n\}$ obeying the relation
 $$
 x_{n+1}+cx_{n}+x_{n-1}=3x_{n}^3, \eqno(1.1)
 $$
 which has been studied in \cite{Berg1,Berg2}.
 The coefficient $3$ can be re-scaled to an arbitrary
 positive number, so only $c$ is the
 parameter of dynamical significance.
 Let $(a_n,b_n)=(x_n,x_{n-1})$, Eq.(1) can be rewritten as a
 two-dimensional conservative map,
 $$
 (a_{n+1},b_{n+1})=F(a_{n},b_{n})=(3a^3_n-ca_n-b_n,a_n).
 \eqno(1.2)
 $$
 The system contains a fixed point $O=(0,0)$, which is unstable
 when $|c|>2$. As in Ref. \cite{Berg2}, we consider the
 orbits that are homoclinic to point $O$ at $c>2$.
 The paper is organized as follows. Section 2 gives a preliminary
 discussion of the map. Section 3 contains a general description
 of its SD. In Sec. 4, symmetric HOs are discussed in detail. This
 is followed by a brief summary.

 \section {Preliminary}
 \subsection{Symmetries}
  Map (1.2) is invariant under the spatial reversion
  $(a,b)\rightarrow (-a,-b)$. Moreover, it has two
  inverse symmetries, $I_1: (a,b) \rightarrow (b,a)$
  and $I_2: (a,b) \rightarrow (-b,-a)$. $I_1$ and $I_2$
  satisfy $I_1 FI_1=I_2FI_2 =F^{-1}$ and $I_1^2=I_2^2=I$.
  The inverse symmetries determine four fundamental
  symmetry lines,  \\
  $$
   \begin{array}{clcl}
  sl_1: &I_1(a,b)=F(a,b) & {\rm or} &3a^3-ca-2b=0 \\
  sl_2: &I_1(a,b)=(a,b)  &  &a-b=0 \\
  sl_3: &I_2(a,b)=F(a,b) &  &a=0 \\
  sl_4: &I_2(a,b)=(a,b)  &  &a+b=0 \
  \end{array}
  \eqno(2.1)
  $$
 For points on these lines, their forward and backward trajectories
 are related by simple rules.
 Specifically, $F^k(X)$ coincides with
 $I_1F^{1-k}(X)$,$I_1F^{-k}(X)$,$I_2F^{1-k}(X)$ or $I_2F^{-k}(X)$
 if $X$ belongs to $sl_1$,$sl_2$,$sl_3$ or $sl_4$,
 respectively. Therefore, if $X$ further belongs to
 the unstable manifold of a fixed point, it must be a homoclinic
 or herteroclinic point. This property
 was used to locate HOs with symmetry \cite{Berg2}.

 \subsection {Normal form for stable and unstable manifolds}

 Numerical investigation of HOs relies essentially on a
 precise computation of the stable and unstable
 manifolds. For canonic systems, the invariant manifold
 of a unstable fixed point $X_0=(a_0,b_0)$
 can be determined by the Birkhoff normal form\cite{ozorio}.
 The manifold is given by the parameterized curve
 $X(t)=(a(t),b(t))$ with
 $$
 a(t)=\sum_{i=0}^\infty a_it^i {\qquad\rm and\qquad}
 b(t)=\sum_{i=0}^\infty b_it^i,
 \eqno(2.2)
 $$
 where coefficients $\{a_i, b_i\}$ are determined by requiring
 $$
 X(\lambda t)=F(X(t)).
 \eqno(2.3)
 $$
 Here $\lambda$ is one of the eigenvalues of
 $F$ at $X_0$, specifically, $|\lambda|>1$ for the unstable
 manifold and $|\lambda|<1$ for the stable one. Take
 $X_0=O$ as an example. Its unstable manifold can be
 written as
 $$
 O_{u}(t)=(G(\lambda t),G(t)),
 \eqno(2.4)
 $$
 where $\lambda=-(c+\sqrt{c^2-4})/2$ and $G(t)=g_1t+g_2t^2+g_3t^3...$
 with $g_1=1$ and
 $$g_k=\frac{3}{\lambda^k+\lambda^{-k}+c}\sum_{q+r+s=k}g_qg_rg_s
 \eqno(2.5)
 $$
 for $k>1$. When $k\rightarrow \infty$, $g_k$ drops very fast
 if $c$ is not too small (e.g. $c>3$). In our numerical study,
 $O_u(t)$ at $t \sim 1$ is computed by the series expansion with
 a reasonable cutoff. Applying $F$ on a precisely determined
 piece of $O_u(t)$ (2.4), we may obtain $O_u(t)$ at large $t$
 with high precision.

 \section {Symbolic dynamics: general description}
 \subsection {Fully-developed horse-shoe and complete
  symbolic dynamics}
  The global dynamics of (1.2) at large $c$ is most
  conveniently accounted for based on the relation
  between the stable and unstable manifolds of
  the fixed points
  $A=(x_d,x_d)$ and $B=(-x_d,-x_d)$, where $x_d=\sqrt{(c+2)/3}$.
  Figure 1a shows these invariant manifolds at $c=5.7$, denoted by
  $A_s,A_u,B_s$ and $ B_u$, respectively                                                                                                              .
  The first part of $A_s$,
  $\overline {ACE}$, is approximately a horizontal
  line. $F^{-1}$ maps $\overline {AC}$
  to $\overline {ACE}$ and two additional nearly
  horizontal segments, $\overline {EG}$ and $\overline {GJ}$,
  forming a s-turn. The first three parts of
  $A_u$, $B_s$ and $B_u$ can be generated from $A_s$
  by taking advantage of the symmetry.

  Under the action of $F^{-1}$,
  the area confined by the first segments of $A_s$, $A_u$,
  $B_s$ and $B_u$ (the diamond-shaped $ACBD$ in Fig. 1a) is
  transformed to the s-shaped region $ACEGJBDFHK$, indicating that
  the global dynamics exhibits a well-developed
  horse-shoe. This is schematically shown in Fig.1b.
  After taking into consideration of the symmetry,
  we can see that this geometric structure
  is guaranteed by the intersection between $\overline {EGJ}$
  consisting of the second and third segments of $A_s$ and
  $\overline {AD}$, the first part of $A_u$.
  The points of intersection correspond
  to two orbits that are homoclinic to point $A$.
  In fact, the two points lie on $sl_1$ and hence
  correspond to HOs with reflection symmetry. With the decrease
  of $c$, the two points move toward each other and collide
  at $c=c^{\ast}\approx 5.453254835$, leading a tangency
  of $A_s$, $A_u$ and $sl_1$.

  As a prototype of chaotic dynamics, the global
  behavior of horse-shoe is well-known.
  In our case, it is conveniently described by
  a complete ternary SD.

   {\sl Symbolic representation of points} \qquad
  All stable (or unstable) manifolds, that generally refer to invariant
  vector field and shrink under the action of $F$
  (or $F^{-1}$)\cite{Zheng}, can be naturally
  divided into three groups and coded
  with $+$, $0$ and $-$, respectively (see Fig. 1).
  Correspondingly, a point $X$ in phase
  space is represented by a bilateral symbolic sequence
  $$
  S(X)=...s_{-2}s_{-1}\bullet s_0 s_1 s_2...
  \equiv S_{-}(X)\bullet S_{+}(X),
  \eqno (3.1)
  $$
  where $s_k\in \{+,0,-\}$
  is the code for the stable (or unstable)
  manifold on which $F^{k}(X)$ (or $F^{k+1}(X)$ )
  is located.  For example, the symbolic sequence for points
  $C$, $J$, $D$ and $K$ in Fig.1a are
  $-^{\infty}\bullet +^{\infty}$,
  $-^{\infty}\bullet -+^{\infty}$,
  $+^{\infty}\bullet -^{\infty}$
  and $+^{\infty}\bullet + -^{\infty}$,  respectively.

  The completeness of
  SD means that each sequence is realized by at least
  one phase space point. By definition we have
  $$
  S \circ F (X) = {\cal F} \circ S (X),
  \eqno (3.2)
  $$
  where $\cal {F}$ is the right shift of $``\bullet"$ with respect
  to the sequence. Therefore, a bilateral symbolic sequence
  without $``\bullet"$ can be used to represent an orbit
  without its initial point specified. For example,
  the symbolic sequences of the fixed points $A$, $B$ and $O$
  are $+^{\infty}$,$-^{\infty}$ and $0^{\infty}$, respectively.

  {\sl Order of symbolic sequences} \qquad
  A point is indicated by its forward sequence $S_+$
 (hence stable manifold) and backward sequence $S_-$
 (unstable manifold). The location of
  stable manifolds in the phase space induce a nature order to
  forward symbolic sequences,
  $\bullet s_0 s_1 s_2... > \bullet s'_0 s'_1 s'_2...$
  if the stable manifold of $\bullet s_0 s_1 s_2...$
  is above that of  $\bullet s'_0 s'_1 s'_2...$.
  The assignment of letters implies
  $$
  \bullet + > \bullet 0 > \bullet -,
  \eqno (3.3)
  $$
  i.e., any sequence beginning with $``+"$ is larger than
  any sequence beginning with $``0"$, which
  is in turn larger than any sequence beginning with $``-"$.
  The comparison of two arbitrary sequences
  relies on the fact that $+$ and $-$ preserve the
  order while $0$ reverses it. (Note that the eigenvalues of
  fixed points $A$ and $B$ are positive while those of point $O$
  are negative.) Assume that $S=s_1s_2...s_ns_{n+1}...$ and
  $S'=s_1s_2...s_n s'_{n+1}...$ and $\bullet s_{n+1} >\bullet s'_{n+1}$,
  then $\bullet S > \bullet S'$
  (or $\bullet S < \bullet S'$)
  if the number of $``0"$ in
  $s_1s_2...s_n$ is even (or odd).
  For example, the first three segments of $A_s$ correspond
  to $\bullet +^{\infty}$, $\bullet 0+^{\infty}$
  and $\bullet -+^{\infty}$, respectively.
  It can be easily verified that
  the first among the three is the largest forward
  sequence, and that there is no sequence which is smaller
  than the second and at the same time larger than the third.
  Similarly, backward sequences can be
  ordered in accordance with the location of their unstable
  manifolds in the phase space. We have
  $$
  ...s_3s_2s_1\bullet > ...s'_3s'_2s'_1 \bullet
  \Leftrightarrow
  \bullet s_1s_2s_3... > \bullet s'_1s'_2s'_3... .
  \eqno (3.4)
  $$
 We shall indicate a point, an
 orbit or a piece of invariant manifold with the same symbolic
 sequence if there is no danger of causing confusion.

 \subsection {Under-developed horse-shoe and truncated symbolic dynamics}
  When $c<c^{\ast}$, the second and third segments of
  $A_s$ do not cross the first segment of $A_u$.
  In this case, there are some ``forbidden" sequences,
  e.g. $+^{\infty}0+^{\infty}$
  and $+^{\infty}-+^{\infty}$,
  which cannot be realized by any real orbits.
  Consider a sequence $S=S_{-}S_{+}$.
  When $c>c^{\ast}$, it corresponds to
  a real orbit on which there is a point being
  the intersection of stable manifold $\bullet S_+$
  and unstable manifold $S_- \bullet$.
  When we reduce $c$, the two segmens of manifold will vary
  continuously with $c$ and they must be tangent
  at a critical value of $c=c_b$ before $S$ is forbidden.
  The tangency can be viewed as a collision of two
  orbits. Since the two colliding orbits are infinitely
  close when $c\rightarrow c_b$, generally their
  corresponding sequences differ by only one symbol, say,
  one is $S_1 0S_2$ and the other is
  $S_1-S_2$. (A logically more meaningful statement
  is that, to maintain the simple order of invariant
  manifolds, the partition line that separates
  $S_1 0\bullet$ and $S_1 -\bullet$ should
  pass the tangent point\cite{Zheng}).
  The tangency of stable manifold $\bullet S_2$  with
  unstable manifolds $S_1 0\bullet$ and $ S_1 -\bullet $
  implies a forbidden zone in the space of
  sequences or symbolic space\cite{Zheng}. Specifically speaking,
  sequence $S'=S'_1 \bullet S'_2$ with
  $$
  S_1- \bullet \leq S'_1 \bullet \leq S_1 0 \bullet
  {\rm \quad ~and\quad ~}
  \bullet S'_2 \ge \bullet S_2
  \eqno(3.5a)
  $$
  corresponds to no point in the phase space (Fig.2a).
  Similarly, if the two colliding orbits are
  $S_1 0S_2$ and $S_1 +S_2$, the forbidden zone is given by
  $$
   S_1+ \bullet \geq S'_1 \bullet \geq S_1 0 \bullet
  {\rm \quad ~and\quad ~}
  \bullet S'_2 \leq \bullet S_2.
  \eqno(3.5b)
  $$
  Of course, any sequence $S'$ with certain
  ${\cal F}^k (S')$ satisfies
  (3.5) is also forbidden.

  Due to symmetry, it is not exceptional in our case
  that the sequences of the two colliding orbits differ
  by two symbols. The forbidden zone due to
  such collision will be also discussed.

 \section {Symmetric homoclinic orbits}
  We now focus on the orbits that are homoclinic to the fixed
  point $O$. The unstable manifold of $O$ is given by (2.4).
  From the symmetry, the stable manifold can be
  written as $O_s(t)=(G(t),G(\lambda t))$. By definition,
  HOs correspond to the solutions of
  equations $G(\lambda t)=G(t')$ and $G(t)=G(\lambda t')$.
  However, the equations have little practical use
  because $G(t)$ changes violently at large $t$.
  The more useful knowledge is the global location of
  the stable and unstable manifolds.
  When $c>c^{\ast}$, each sequence $0^{\infty}S0^{\infty}$
  corresponds to a HO. To locate the orbit in the phase space,
  we can scan initial point $X$ on $0^{\infty} \bullet$,
  the first segment of $O_u$, searching for the target
  $S_+(X)=S0^{\infty}$.  For a symmetric HO,
  the target can be replaced by $F^{k}(X)\in sl_i$
  or alternatively the scan can be performed on a symmetry line.
  In all cases, the order of forward sequences
  can be used for an effective searching strategy.
  Having a HO at large $c$, we can trace it to
  small $c$ until it collides with another HO
  and disappears. We shall not go into details
  about the locating of HOs.
  We shall focus on the compatibility between HOs
  at $c<c^{\ast}$.

   In the following discussion, the stable and unstable
   manifolds are always referred to as $O_s$ and $O_u$.
   We first give another description of the
   forbidden zone induced from the tangency of
   stable and unstable manifolds, or the collision of two HOs.
   Let $X$ and $X'$ be two homoclinic points which
   collide each other when $c\rightarrow c_b$.
   For $c>c^{\ast}$, the stable and unstable manifolds
   connecting  $X$ and $X'$ form a closed curve
   ${\cal O} (X,X')$ (see Fig.~2), which shrinks to a
   single point when $c=c_b$. Note that both $O_s$ and
   $O_u$ are continuous and self-avoiding curves.
   Consider a homoclinic point enclosed by ${\cal O} (X,X')$
   or lying on ${\cal O} (X,X')$. Being the
   intersection  of stable and unstable manifolds,
   the point cannot cross ${\cal O}(X,X')$ when $c$ is
   continuously varied.
   The area enclosed by ${\cal O}(X,X')$ at
   $c>c^{\ast}$ vanish before $c=c_b$.
   Therefore, ${\cal O}(X,X')$ defines the
   forbidden zone corresponds to the collision of
   $X$ and $X'$.

   It can be seen that the forbidden zone defined
   by ${\cal O}(X,X')$ coincides with that defined by
   (3.5) if $S(X)$ and $S(X')$ differ by
   only one symbol (see Fig.2b). To study the
   evolution of symmetric HOs, we need also consider the
   the case that $S(X)$ and $S(X')$ differ by two symbols.
   For this purpose, the structure of stable and unstable
   manifolds should be studied in detail.
   Consider the $3^k$ segments of stable manifold of the form
   $\bullet S0^{\infty}$, where $S$ exhausts all the
   ternary strings with length $|S|=k$.
   The relative locations of these segments of
   manifold in the phase space are
   determined by the order of $\bullet S$.
   We ask that how the $3^k$ segments are
   connected to form a continuous curve.
   All these segments are generated from
   $\bullet 0^{\infty}$ by $F^{-k}$,
   i.e., ${\cal F}^{-k}(S \bullet 0^{\infty})
   =\bullet S0^{\infty}$.
   We may cut the segment $\bullet 0^{\infty}$,
   according to backward sequences, into $3^k$ pieces,
   each of which are coded with $S \bullet$.
   The way to joint these $3^k$ pieces into the whole segment
   $\bullet 0^{\infty}$ is determined by the order of the
   $3^k$ strings $S \bullet$.
   After expanding under $F^{-k}$, the segment $\bullet 0^{\infty}$
   becomes the $3^k$ segments of $\bullet S0^{\infty}$, so
   they should be connected also according to $S \bullet$.
   In the simplest case of $k=1$, three segments of $O_s$
   from top to bottom are $\bullet +0^{\infty}$,
   $\bullet 0^{\infty}$ and $\bullet -0^{\infty}$.
   Forming an s-turn, $\bullet +0^{\infty}$ and
   $\bullet -0^{\infty}$ joint with
   $\bullet 0^{\infty}$ at its left and right ends, respectively.
   For $k=2$, the nine
   segments of $O_s$ are ordered from top to bottom as
   $$
   ++,\quad +0,\quad +-,\quad 0-,\quad 00,\quad 0+,\quad -+,
  \quad -0\quad {\rm and}\quad --,
   $$
  as shown in Fig.~3 schematically by horizontal lines.
  In the figure $S$ is also taken as the abbreviation for
  $\bullet S 0^{\infty}$.
  These segments are connected end to end according to
  the following order of backward $S\bullet$:
  $$
  ++,\quad 0+,\quad -+,\quad -0,\quad 00,\quad +0,
 \quad +-,\quad 0-\quad {\rm and}\quad --.
  $$
  The arrangement is shown in Fig.3.
  We see that pair $(00, +0)$ joint at their left ends,
  $(+0,+-)$ joint at their right ends,
  $(+-,0-)$ joint at their left ends, and so on.
  In general, the sequence pair of connected
  segments of stable manifold take one of the following forms,
  $$
 \left \{ \begin {array}{l}
 \rm{(i)}\quad ~S=-^{m}0S_0 \quad {\rm and} \quad S'=-^{m}+S_0,\\
 \rm{(ii)}\quad S=+^{m}0S_0 \quad {\rm and} \quad S'=+^{m}-S_0,\\
 \end {array}
 \right.
  $$
  where $m\geq 0$. Segment pair of type (i)
  are connected at their left ends
  while pair of type (ii)
  are connected at their right ends.
  By taking advantage of the symmetry, a
  parallel description can be given to the unstable
  manifold.

  Now we study the forbidden zone implied by
  the collision of two symmetric HOs.
  Let us first consider two examples.
  In the first case, $S(X)=0^{\infty}\bullet
  +++0^{\infty}$ and $S(X')=0^{\infty}-\bullet +++-0^{\infty}$.
  From the above discussion, $0^{\infty}\bullet$
  and $0^{\infty}-\bullet$ are connected at
  their upper ends while
  $\bullet +++0^{\infty}$ and $\bullet +++-0^{\infty}$
  are connected at their right ends. The forbidden zone
  enclosed by ${\cal O}(X,X')$ is shown in Fig.~2c.
  We see that it is the union of two
  forbidden zones, one is defined by ${\cal O}(X,X'')$
  and the other by ${\cal O}(X',X'')$, where
  $S(X'')=0^{\infty}\bullet +++-0^{\infty}$
  (or $0^{\infty}-\bullet +++0^{\infty}$).

  The next example is
  $S(X)=0^{\infty}-0\bullet 00-0^{\infty}$
  and $S(X)=0^{\infty}-+\bullet 0+-^{\infty}$.
  Now $0^{\infty}-0\bullet$ and
  $0^{\infty}-+\bullet$ are connected at
  the bottom. On the other hand,
  $\bullet 00-0^{\infty}$ and
  $\bullet 0+-0^{\infty}$ are
  not joint neighboring segments on $O_s$.
  They are connected by
  $\bullet -0-0^{\infty}$ and
  $\bullet -+-0^{\infty}$. The area enclosed
  by ${\cal O}(X,X')$ is shown in Fig. 2d, which
  again turns out to be the union of two
  regions enclosed by
  ${\cal O}(X,X'')$ and ${\cal O}(X',X'')$ with
  $S(X'')=0^{\infty}-0\bullet 0+-0^{\infty}$
  (or $0^{\infty}-+\bullet 00-0^{\infty}$).

  In general, suppose $S(X)=0^{\infty}S_10\bullet
  S_2 0S_3 0^{\infty}$
  and $S(X')=0^{\infty}S_1 \sigma_1\bullet
  S_2 \sigma_2 S_3 0^{\infty}$, where $\sigma_{1,2}\in \{+,-\}$.
  This covers all the colliding symmetric HOs with
  two different letters. As seen in the two examples,
  it can be shown that the forbidden zone enclosed
  by ${\cal O}(X,X')$ is the union of that
  defined by ${\cal O}(X,X'')$ and ${\cal O}(X',X'')$,
  where $S(X'')=0^{\infty}S_1 0\bullet
  S_2 \sigma_2 S_3 0^{\infty}$
  (or $0^{\infty}S_1 \sigma_1\bullet
  S_2 0 S_3 0^{\infty}$).
  In other words, when forming the forbidden zone,
  the collision of $X$ and $X'$ can be regarded as
  the triple collision involving $X$, $X'$ and $X''$.
  Since either the pair of $S(X)$ and $S(X'')$ or
  the pair of $S(X')$ and $S(X'')$
  differ by one symbol, the corresponding forbidden
  zone is given by (3.5). Thus, the rule can be used to
  construct the forbidden zone for the above general case.

  We have numerically calculated all symmetric HOs
  $0^{\infty}S0^{\infty}$ with $|S|\leq 7$ and traced
  them from $c=6$ to collision parameters $c_b$.
  Among the $160$ orbits, only
  $0^{\infty} \pm 0^{\infty}$ and
  $0^{\infty} \pm \mp 0^{\infty}$ persist till $c=2$.
  We obtained $94$ collisions, each of them corresponds to
  a tangency of $O_s$ and a symmetry line (see Fig.4 for
  some examples).
  Due to symmetry, only half of the collisions are
  independent. They are ordered with descending $c_b$ and
  listed in Tab.~1.  We further examine the compatibility
  between symmetric HOs. Collision of a
  pair of symmetric HOs may implies the
  elimination of some others. For example,
  it can be verified that the forbidden zone
  of the colliding HO pair $(S,S')=(+-+-,+00-)$
  includes the HO sequence pair $(-+-,-0-)$.
  Therefore, the collision  of HO pair $(-+-,-0-)$
  should occurs at a larger $c$ than that for $(+-+-,+00-)$.
  In fact, the former appears as the 9-th collision in
  Tab. 1 while the latter is the 16-th.
  We find that all such restrictions
  in the 47 pairs of sequences in Tab.1
  are in accordance with the order of $c_b$ (Fig.5).

 \section {Summary}
  In the above we have discussed the symbolic dynamics of
  system (1.2) with a focus on the orbits that are homoclinic
  to point $O=(0,0)$. When $c>c^{\ast}$, the global dynamics
  is described by a complete ternary symbolic dynamics, so that
  each sequence $0^\infty S 0^{\infty}$ corresponds a homoclinic
  orbit. The location of these orbits are qualitatively
  determined by their corresponding sequences according to
  simple rules. With the decrease of $c$, more and more
  symmetric homoclinic orbits collide in pair and disappear.
  Based on the geometrical constraint that any homoclinic point
  cannot cross a region enclosed by the stable and unstable
  manifolds, the compatibility between homoclinic orbits
  is discussed.

\newpage
\parindent -5mm
\begin{center}
Figure Captions
\end{center}

 Fig.1 (a) First three segments of $A_s$, $A_u$, $B_s$ and
       $B_u$ at $c=5.7$. The dotted lines ($a,b=\pm \sqrt{c}/3$)
       give a convenient partition of manifolds. (b)Schematic
       diagram of hose-shoe. The square $ACBD$ is
       mapped by $F^{-1}$ to the shaded area.

 Fig.2 Schematic diagrams to show forbidden sequences
        due to tangency between the stable and unstable
        manifolds. Figure (a) explains rule (3.5a).
        The shadowed areas in (b-d) represent
        the forbidden zones when $X$ and $X'$ collide.

 Fig.3 Arrangement of the first $9$ segments of $O_s$.

 Fig.4 Tipical tangencys of $O_u$ (thick line) and symmetry
      line (thin line). Each tangeny represents a collision
      of symmetric homoclinic orbit pair.

 Fig.5 Compatibility between symmetric homoclinic
       orbit pairs listed in Tab.1. A dot at $(i,j)$ means
       that the collision of $i$-th pair implies
       the elimination the $j$-th pair. Note that all dots
       are located within the region $i>j$.

\newpage

 {\tiny

 \begin {center} {Tab.1 Critical parameter of symmetric HOs}

 \end {center}

 $$
 \begin{array}{ccll|ccll}
 \hline
\quad {\rm order} \quad  &  c_b &\quad \quad S &\quad \quad S' &
\quad {\rm order} \quad  &  c_b &\quad \quad S &\quad \quad S'
  \\ \hline
  1 &    5.4510677& +++-+++  &  +++0+++ &
  25&    3.3139443&  ++0--   &  -++0--+ \\
  2 &   5.4158176&  ++-++   &   ++0++  &
 26 &    3.3129838& ++0-0++  & -++0-0++-\\
  3 &   5.3937939& +--0--+  &  +--+--+ &
27  &    3.3064455& ++00++   & -++00++- \\
  4 &   5.1064013& +-+++-+  &  +0+++0+ &
28  &    3.3029884& ++000--  & -++000--+\\
  5 &   5.0779902& +-++-+   &  +0++0+  &
29  &    3.2978459&   ++     &   -++-   \\
 6  &   4.9161551& ++-+-++  &  ++0+0++ &
30  &    3.2907686& ++-0+--  & -++-0+--+\\
 7  &   4.8549163& +0-+-0+  &  +0-0-0+ &
31  &    3.2821047& ++--++   & -++--++- \\
 8  &   4.7672774& +-+0-+-  &  +0+0-0- &
32  &    3.2817081&  ++--    &  -++--+  \\
9   &   4.6777145&   +-+    &    +0+   &
33  &    3.2807320& ++---++  & -++---++-\\
10  &    4.5252523& +-+0+-+  &  +0+0+0+ &
34  &    3.1683361&  +-0-+   &   +000+  \\
11  &    4.4437742&  +-+-+   &   +0+0+  &
35  &    2.9640007& +0---0+  & -+0---0+-\\
12  &    4.3722249& +-+-+-+  &  +0+-+0+ &
36  &    2.9619403& +0--0+   & -+0--0+- \\
13  &    4.2904264& +-+-+-   &  +0+-0-  &
37  &    2.9381982& +0-0+0-  & -+0-0+0-+\\
14  &    4.1671612& ++-+--   &  ++00--  &
38  &    2.9373534&  +0-0+   &  -+0-0+- \\
15  &    3.6530814& ++-0-++  &  ++000++ &
39  &    2.9205856&   +0-    &   -+0-+  \\
16  &    3.6437729&  +-+-    &   +00-   &
40  &    2.9083662& +0-+0-   & -+0-+0-+ \\
17  &    3.5589054& +-0-0-+  &  +00-00+ &
41  &    2.9066760& +-00+-   &  +0000-  \\
18  &    3.3422180& +++++++  & -+++++++-&
42  &    2.7863722& +00+00+  & -+00+00+-\\
19  &    3.3422164& ++++++   & -++++++- &
43  &    2.7742746&  +00+    &  -+00+-  \\
20  &    3.3421968&  +++++   &  -+++++- &
44  &    2.7457804& +-000-+  &  +00000+ \\
21  &    3.3419494&  ++++    &  -++++-  &
45  &    2.6839843&  +000-   &  -+000-+ \\
22  &    3.3400187& +++0---  & -+++0---+&
46  &    2.6170320& +0000+   & -+0000+- \\
23  &    3.3388177&   +++    &   -+++-  &
47  &    2.5628482& +00000-  & -+00000-+\\
24  &    3.3375984& +++---   & -+++---+ \\
 \hline
 \end{array}
$$
 }


\begin{thebibliography}{99}
 \bibitem{Rep} D. Henning and G. Tsironis, Phys. Rep. 307,333
 (1999).
 \bibitem{Fried} N. J. Friedman and E. J. Doedel, SIAM J.
  Numer. Anal. 28, 789 (1991).
 \bibitem{Far}S. Farantos, Comput. Phys. Commun. 108, 240
 (1998).
 \bibitem{Bey1} W. J. Beyn and J. M. Kleinkauf, SIAM J.
  Numer. Anal. 34, 1207 (1997).
 \bibitem{Bey2}W. J. Beyn and J. M. Kleinkauf, Numer.
 Algorithms 14, 25 (1997).
 \bibitem{Berg1} T. Bountis, H. W. Capel, M. Kollmann, J. C. Ross,
  J. M. Bergamin and J. P. Van der Welle, Phys. Lett. A268,50 (2000).
 \bibitem{Berg2} J. M. Bergamin, T. Bountis and C. Jung, J. Phys.
  A: Math. Gen. 33,8059 (2000).
 \bibitem{Zheng} B. L. Hao and W. M. Zheng, Applied Symbolic Dynamics
 and Chaos (Singapor: World Scientific, 1998).
 \bibitem{ozorio} A. M. Ozorio de Almeida, Hamiltonian system: Chaos and
 Quantization (Cambridge Uiversity Press, 1988).



\end{thebibliography}
 
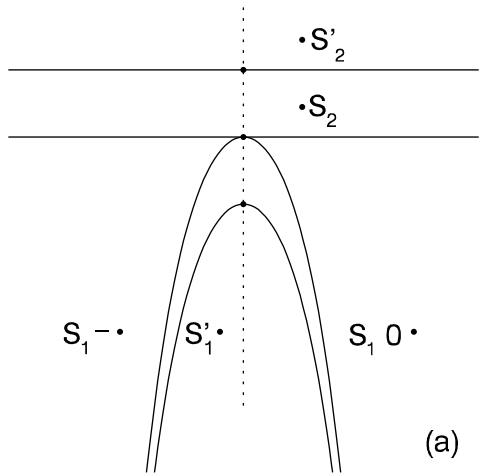
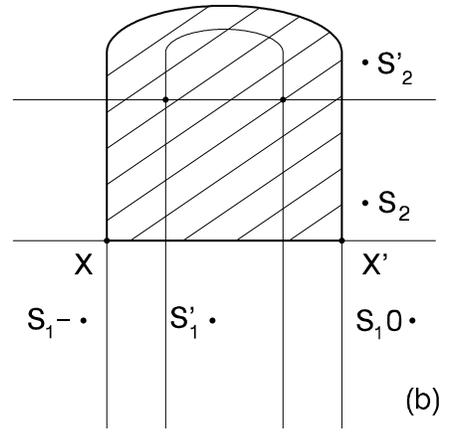
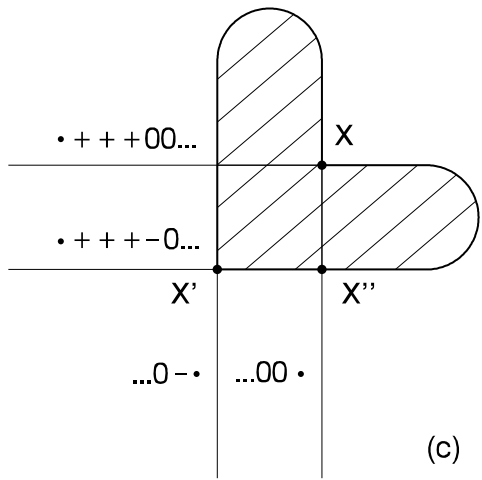
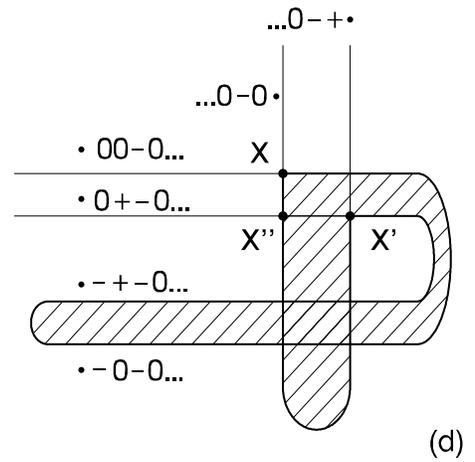
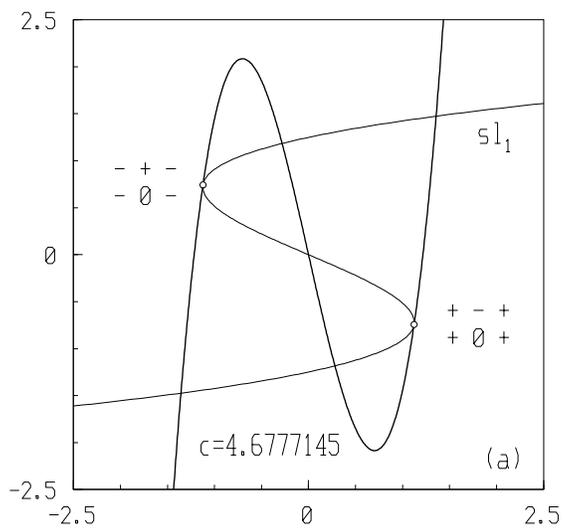
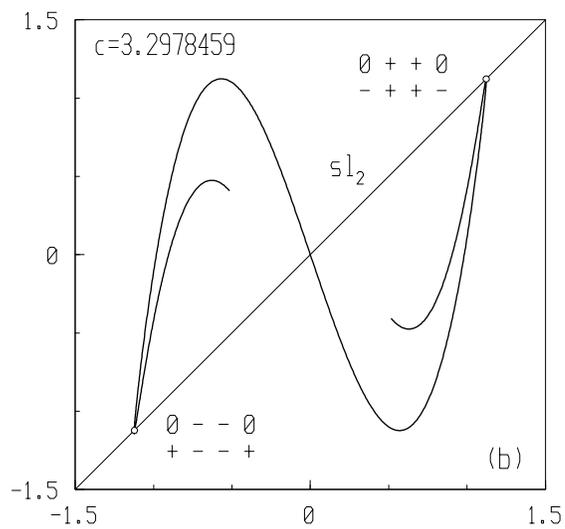
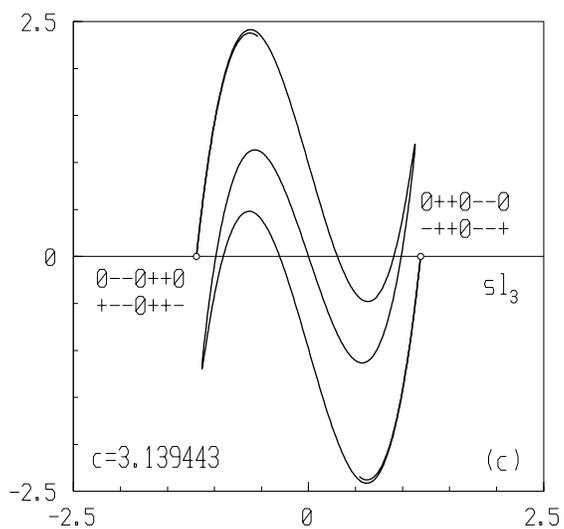
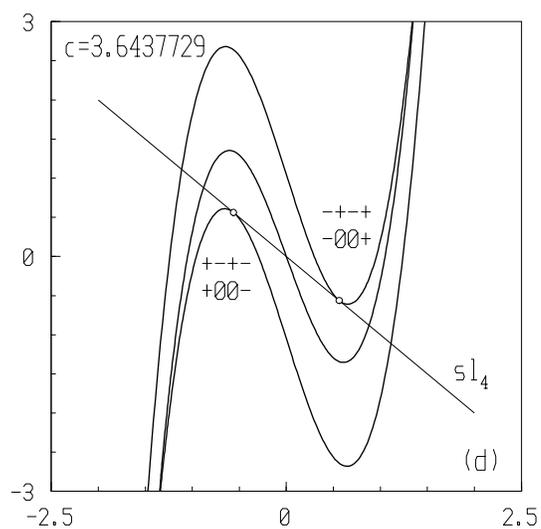
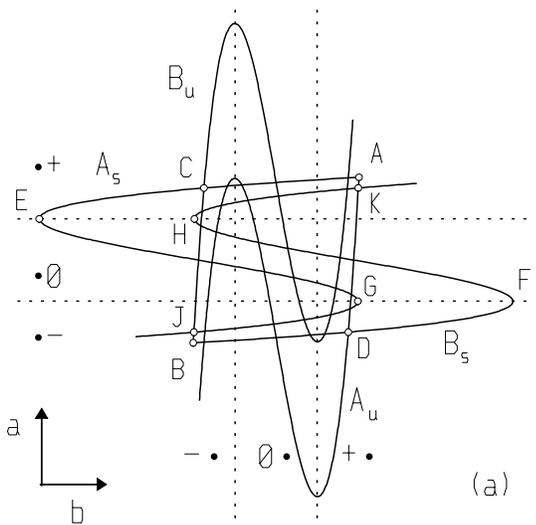
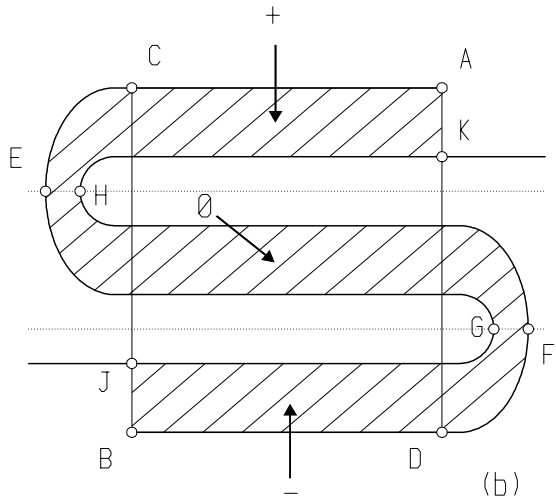
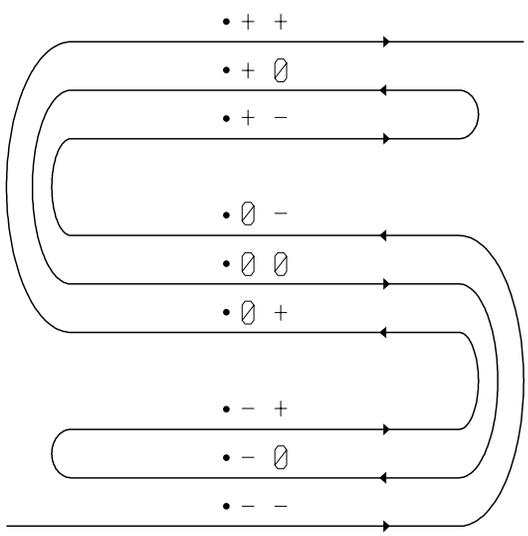
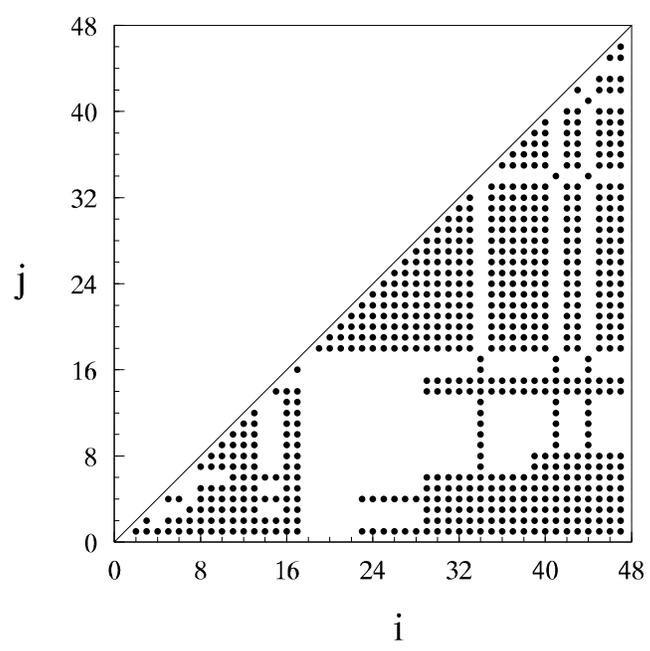
\end{document}